\documentclass{iaus}
\usepackage{graphicx}

\title[Models for Pop I stars] 
{Models for Pop I stars: implications for age determinations}

\author[Georges Meynet et al.]   
{Georges Meynet$^1$, 
Patrick Eggenberger$^2$, Nami Mowlavi$^3$ and Andr\'e Maeder$^1$}

\affiliation{$^1$Geneva University, Geneva Observatory \\
CH-1290 Versoix, Switzerland \\ email: {\tt georges.meynet@unige.ch} \\[\affilskip]
$^2$Institut d'Astrophysique et de G\'eophysique de l'Universit\'e de Li\`ege, All\'ee du 6 Ao\^ut, 17 B-4000 Li\`ege, Belgium {\tt eggenberger@astro.ulg.ac.be} \\[\affilskip]
}

\pubyear{2009}
\volume{258}  
\pagerange{119--126}
\jname{The Age of Stars}
\editors{D. Soderblom et al. eds.}
\begin{document}

\maketitle

\begin{abstract}
Starting from a few topical astrophysical questions which require the knowledge of the age of Pop I stars, we discuss the needed precision on the age in order to make progresses in these areas of research. Then we
review the effects of various inputs of the stellar models on the age determination and try to identify those affecting the most the lifetimes of stars.
\keywords{Physical data and processes: convection, diffusion; Stars: rotation, mass loss, magnetic fields}
\end{abstract}

\firstsection 
              
\section{Importance of reliable age-calibrations for Pop I stars}  


Pop I stars cover a wide range of ages from millions to billions of years and their age determination is useful for studying the evolution of processes with very different timescales,
going from the evolution of planetary systems to the analysis of powerful starbursts in
remote galaxies. A few examples are given below.
\begin{itemize}
\item How does star formation propagate around young star forming regions?
What are the timescales for star formation to trigger star formation (see for instance the very young associations observed in the vicinity of $\eta$Car discussed in the review by Smith \& Brooks 2008)? 
In order to study the time sequence between different associations, relative ages
with an accuracy better than about 10-20\% would be quite useful, that means an absolute accuracy of a few 100 000 years on ages of a few Myr. 
\item What are the ages of young powerful starbursts in remote galaxies?
In distant galaxies, where individual stars cannot be resolved, emission line ratios in the spectrum of the integrated light can be used to determine the age of starbursts (see the review by Leitherer 2005). The typical ages are of a few Myr and the needed precision is of the same order as the one indicated above for the age determination of resolved associations. Let us however stress that here, in addition to the uncertainties pertaining the stellar models, those due to the duration of the burst of star formation, to the Initial Mass function and to the possibility of superpositions of many starbursts makes the exercise still more difficult. 
\item What is the upper mass limit of the White Dwarfs (WD) progenitors? To answer this question one needs
to establish the initial-final mass relationship of WD. This can be done by determining the age of the open clusters where WD are observed. This age is then used, together with the cooling age of the WD, to estimate the mass of the WD progenitor using stellar models. In this process age determination enters in three ways: first through the isochrone fitting of the cluster, second through the cooling age of the WD and finally through an age-mass relation (Weidemann 2000; see also the talks by Kalirai and Richer in this volume). To determine the upper mass limit for the progenitor of WD, accurate determination of the age of clusters with a mass at the turn off around 8 M$_\odot$
are needed, this means ages of the order of a few ten Myr. In this age range, an uncertainty of 20\% on the age translates into an uncertainty of about 1 M$_\odot$ on the mass (thus an uncertainty of $\sim$10\% on the mass).
\item What is the lifetime of a very-hot Jupiter? These giant planets orbit their host stars with orbital periods below 3 days. Given their proximity to their host stars, these planets should undergo some degree of evaporation due to the heating by stellar UV photons. Some models predict a catastrophic destiny for these planets being completely evaporated in a relatively short timescale. If true, very-hot Jupiter could only be observed around the youngest stars. Is it the case?
Melo et al. (2006) find that none of the stars studied in their paper seem to be younger than 0.5 Gyr. Only lower limits for most of the ages of these stars are obtained. To make progresses in this area, restricted ranges of ages for the planet host stars should be obtained.
\item What was the evolution of the chemical gradients in the Milky Way? Nowadays the Milky Way
presents a gradient $\rm d \lg({\rm O/H})/\rm d R$ where $R$ is the galactocentric radius between
-0.07 dex/kpc and -0.04 (see e.g. Daflon \& Cunha 2004). Was this gradient steeper or shallower in the past? From a theoretical point of view the answer is uncertain. In order to answer such a question, one needs objects for which the metallicity and the age can be measured. Open clusters and planetary nebulae are the objects allowing to cover the largest part of the Galaxy history (from a time when its age was about 6 Gyr until today) while simultaneously spanning a wide range of galactocentric radii.
In order to improve our present knowledge of the evolution of this gradient, a precision better
than one half Gyr should be obtained on the ages (precision of about 10\% needed).
\item What is the form of the age-metallicity relation for the thin disc stars of the Galaxy?
For instance Nordstr\"om et al. (2004) have found a small change of the mean metallicity of the thin disk
since its formation and a very substantial scatter in metallicity at all ages. Again an accuracy better than about 10\% on the age would allow to sharpen this picture.
\end{itemize}
While age determinations may reach internal precision sometimes better than 10\%, systematic effects may prevent to reach a similar level of accuracy in absolute ages. 
In the following we shall focus on the systematic effects on age estimates due to uncertainties pertaining various physical effects accounted for in stellar models. 

\section{Comparison of different models for near solar metallicity}

In table 1 below, we list some grids of non-rotating stellar models, covering the case of Pop I stars (the list is not exhaustive and refers to only one paper per considered group!). To make a first comparison, we can plot the relation between the bolometric magnitude at the end of the Main-Sequence phase and the age given by some of these models. This is done in the left panel of Fig.~1. Overall there is a good agreement between the plotted models. The slope of the relation  
[M$_{\rm bol} \propto \sim 15/4 \lg$(age)] well agrees with analytical 
estimates based on the mass-luminosity ($L \propto (\mu^4 M^3)/\kappa$, where $\mu$ is the mean molecular weight, $M$ the total mass and $\kappa$, the mean opacity inside the star) and the mass-age relation
($\tau_{\rm MS}\propto (q X M f c^2)/(\mu^4 M^3 /\kappa)$, where $q$ is the mass fraction of the total mass of the star where nuclear reactions occur, $X$ the mass fraction of hydrogen, $c$ the velocity of light, $f$ the fraction of the initial mass of hydrogen transformed into energy when H-burning occurs, $f$ is equal to 0.007). Thus an error of 0.1 Magnitude implies an error on the age of less than 3\%.

In the right panel of Fig.~1, we indicate the difference between the minimum age and the maximum age given by these models for a given magnitude normalised to the mean age, the mean age being (maximum - minimum age)/2. We see that the age dispersion is of the order of 10\% for ages between 200 Myr and 2 Gyr. 
The dispersion increases up to values between 25 and 30\% for greater and smaller ages. Most of the differences come from different ways of treating the way the overshooting parameter evolves as a function of the initial mass (see the respective references describing the models). 

\begin{table}
\caption{Some grids of non-rotating stellar models for near solar metallicity.}
\begin{center}\scriptsize
\begin{tabular}{lcccc}
 & & & \\
Reference & Masses & $Y$ & $Z$ & $\alpha_{\rm ov}$\\ 
 & & & &\\
Bono et al. 2000$^{(1)}$         & 3 - 15    & 0.27    & 0.02    & 0.0\\
Claret 2004                      & 0.8 - 125 & 0.28    & 0.02    & 0.2\\
Demarque et al. 2004$^{(1)}$     & 4 - 52    & 0.279   & 0.02    & (6)\\
Dominguez et al. 1999            & 1.2 - 9   & 0.28    & 0.02    & 0.0\\
Dotter et al. 2007$^{(1)}$       & 0.1 - 1.8 & 0.274   & 0.0189  & 0.2 see (3)\\
Girardi et al. 2000$^{(1)}$      & 0.15 - 7  & 0.273   & 0.019   & (5)\\
Pietrinferni et al. 2004$^{(1)}$ & 0.5 - 10  & 0.273   & 0.0198  & 0.2 see (3) \& (4)\\
Schaller et al. 1992$^{(1)}$     & 0.8 -120  & 0.30    & 0.02    & 0.2\\
VandenBerg et al. 2006$^{(1)}$   & 0.5 - 2.4 & 0.2715  & 0.0188  & see(2)\\
Ventura et al. 1998              & 0.6 - 15  & 0.274   & 0.017   & FST(7)\\   
 &   & & &\\
\multispan{5}{$^1$ Models for other compositions are available in that paper.\hfill}\\
\multispan{5}{$^2$ Roxburgh criterion calibrated using binaries and clusters.\hfill}\\
\multispan{5}{$^3$ Value adopted in models with well developed convective cores.\hfill}\\
\multispan{5}{        Lower values are used for small convective cores see Dotter et al. (2007)\hfill}\\
\multispan{5}{$^4$ Models without overshooting with the same initial composition are also available.\hfill}\\
\multispan{5}{$^5$ The overshooting is accounted for using the formalism of Bressan et al. (1981).\hfill}\\
\multispan{5}{$^6$ The overshooting is accounted for using the formalism of Demarque et al. (2004).\hfill}\\
\multispan{5}{$^7$ Convection is treated according to the Full Spectrum Turbulence model.\hfill}\\

\end{tabular}
\end{center}
\label{tab1}
\end{table}

\begin{figure}
\includegraphics[width=2.7in,height=2.7in]{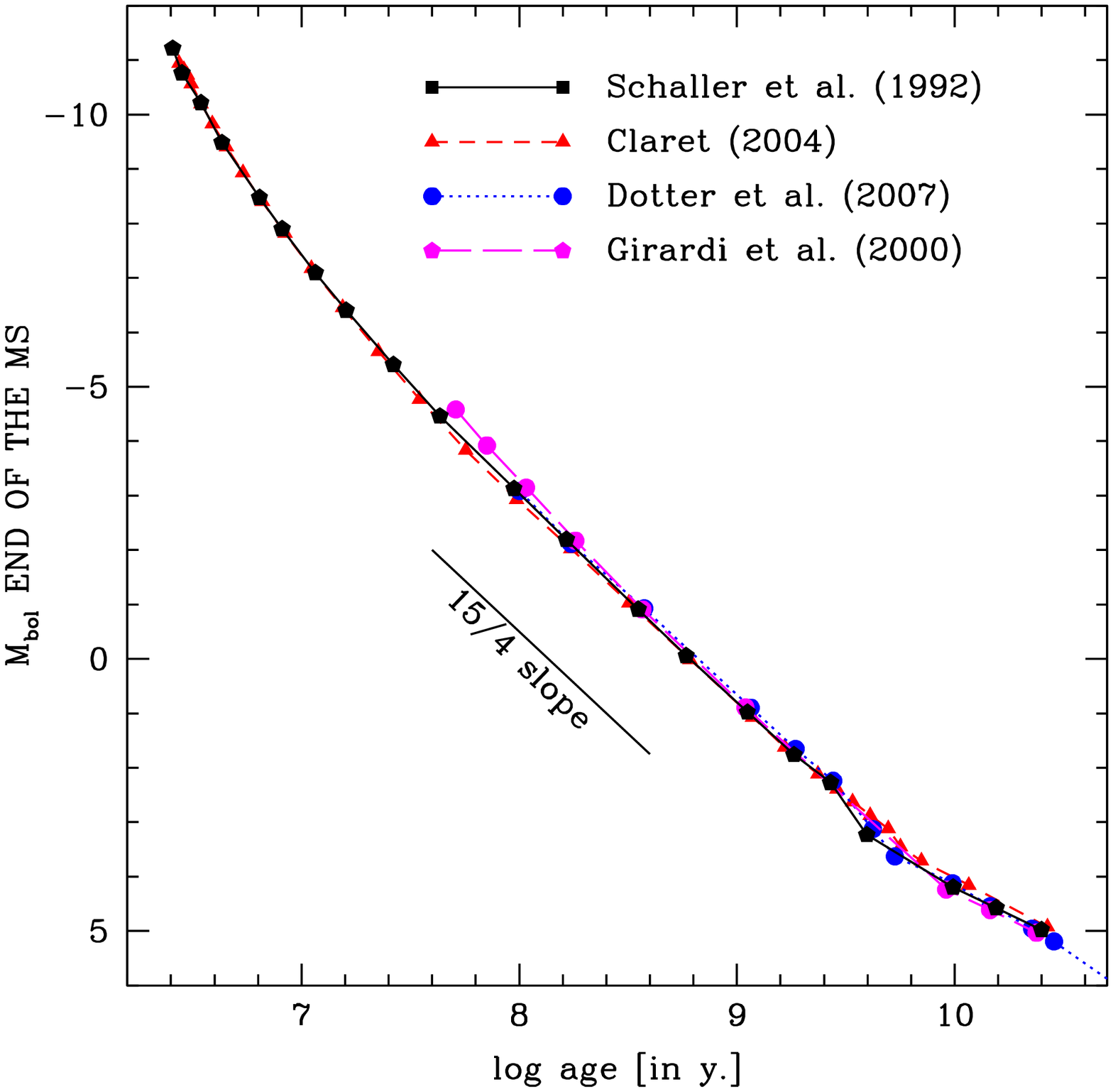}
\hfill
\includegraphics[width=2.7in,height=2.7in]{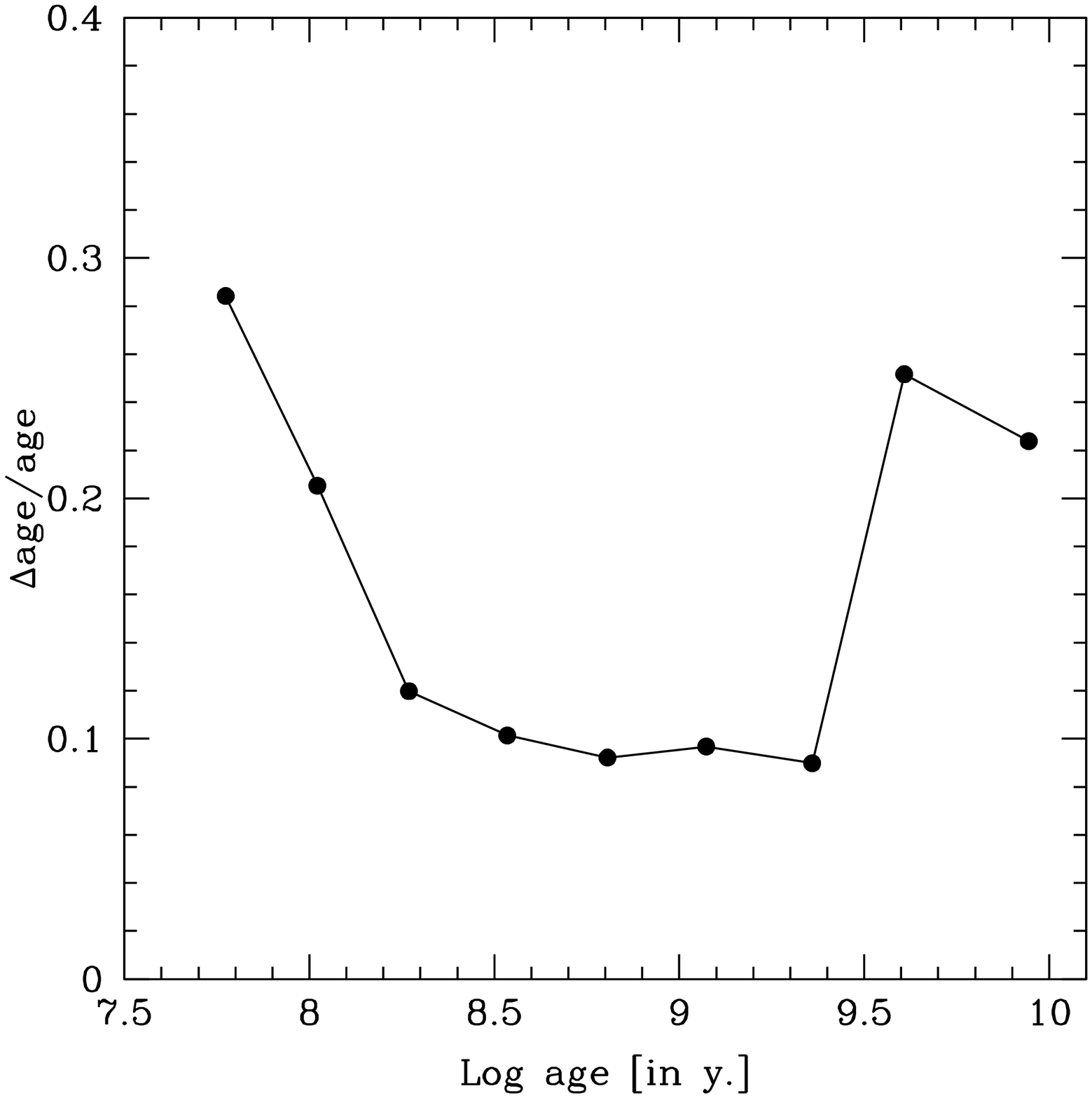}
\caption{{\it Left panels}: bolometric magnitude at the turn off versus the age for different non-rotating stellar models with similar physical ingredients. {\it Right panel}: dispersion of the ages due to the use of different stellar models listed on the left panel.}\label{fig1}
\end{figure}

\section{Effects of a change of the initial abundances}

In non rotating models, the metallicity affects the evolution of stars mainly through its
impact on the radiative opacities, the equation of state and the nuclear
reaction rates. For stars with initial mass greater than about 30 M$_\odot$, mass loss
becomes an important ingredient already during the MS phase and the effects of metallicity
on the mass loss rates have to be taken into account (see the effect on age determination below).
These effects of metallicity on stellar models are discussed in details in Mowlavi et al. (1998).
Metallicity also affects the transport mechanisms induced by rotation (Maeder \& Meynet 2001).
Typically a given initial mass star, starting its evolution with a given initial velocity, will be more efficiently mixed by rotation at low than at high metallicity.

The most important effect of a change of metallicity occurs through the effect on the opacity
(at least for a large range of metallicities and ages).
In general, the opacities increase with increasing metallicity.
This is the case for opacity due to bound-free
and free-free transitions. Using the mass-luminosity relation
seen above, one immediately deduces that the increase
of the opacity produces a decrease of the luminosity of a given initial mass model. 
In contrast, for massive stars free electron scattering is the main opacity source. 
This opacity depends only on $X$
[$\kappa_e\simeq 0.20(1+X)$], which is about constant at
$Z\leq 0.01$ and decreases with increasing $Z$ at $Z \geq 0.01$.
Thus at high $Z$ the luminosity of a massive star increases with the metallicity.

\begin{figure}
\includegraphics[width=2.7in,height=2.7in]{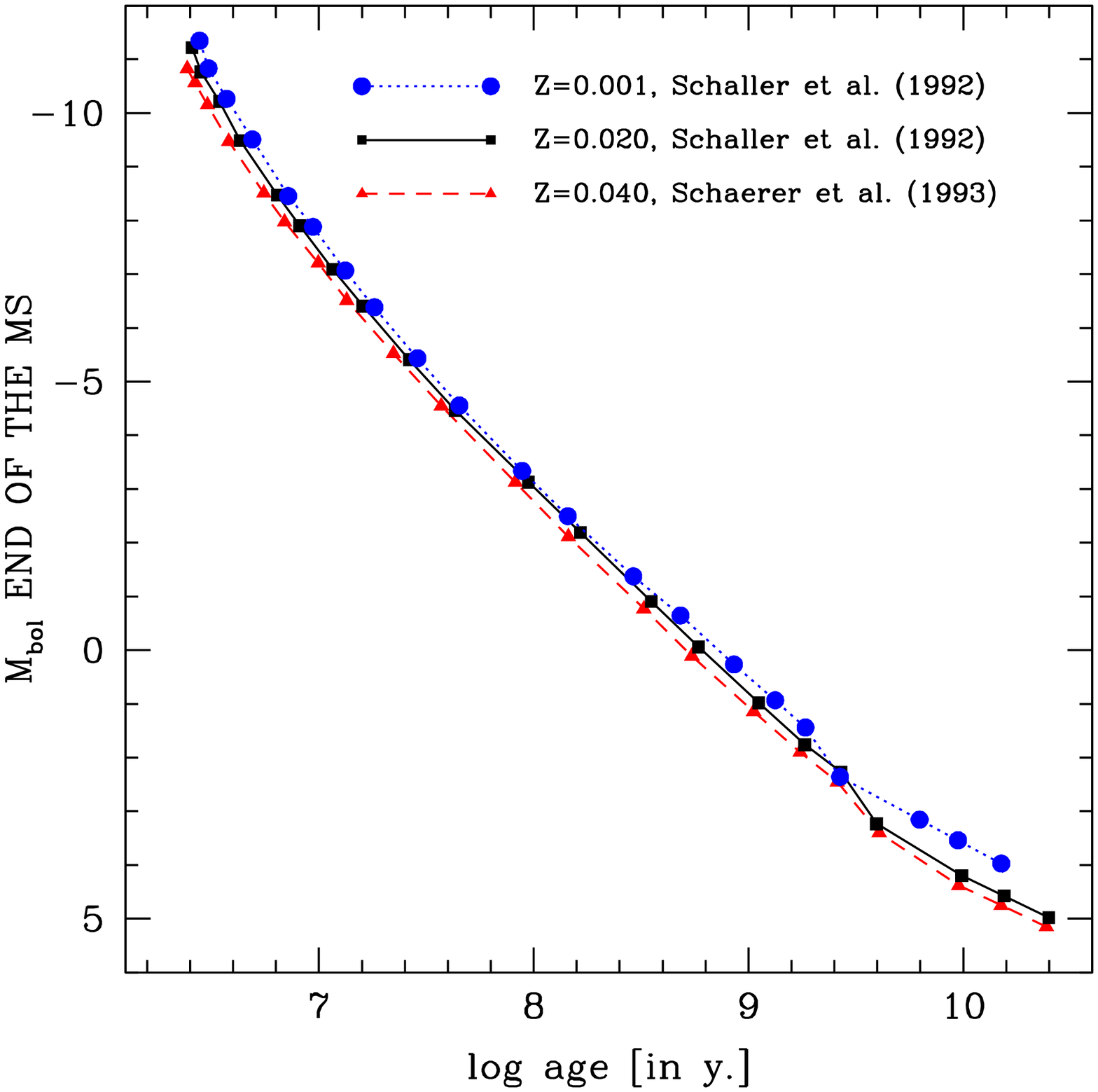}
\hfill
\includegraphics[width=2.7in,height=2.7in]{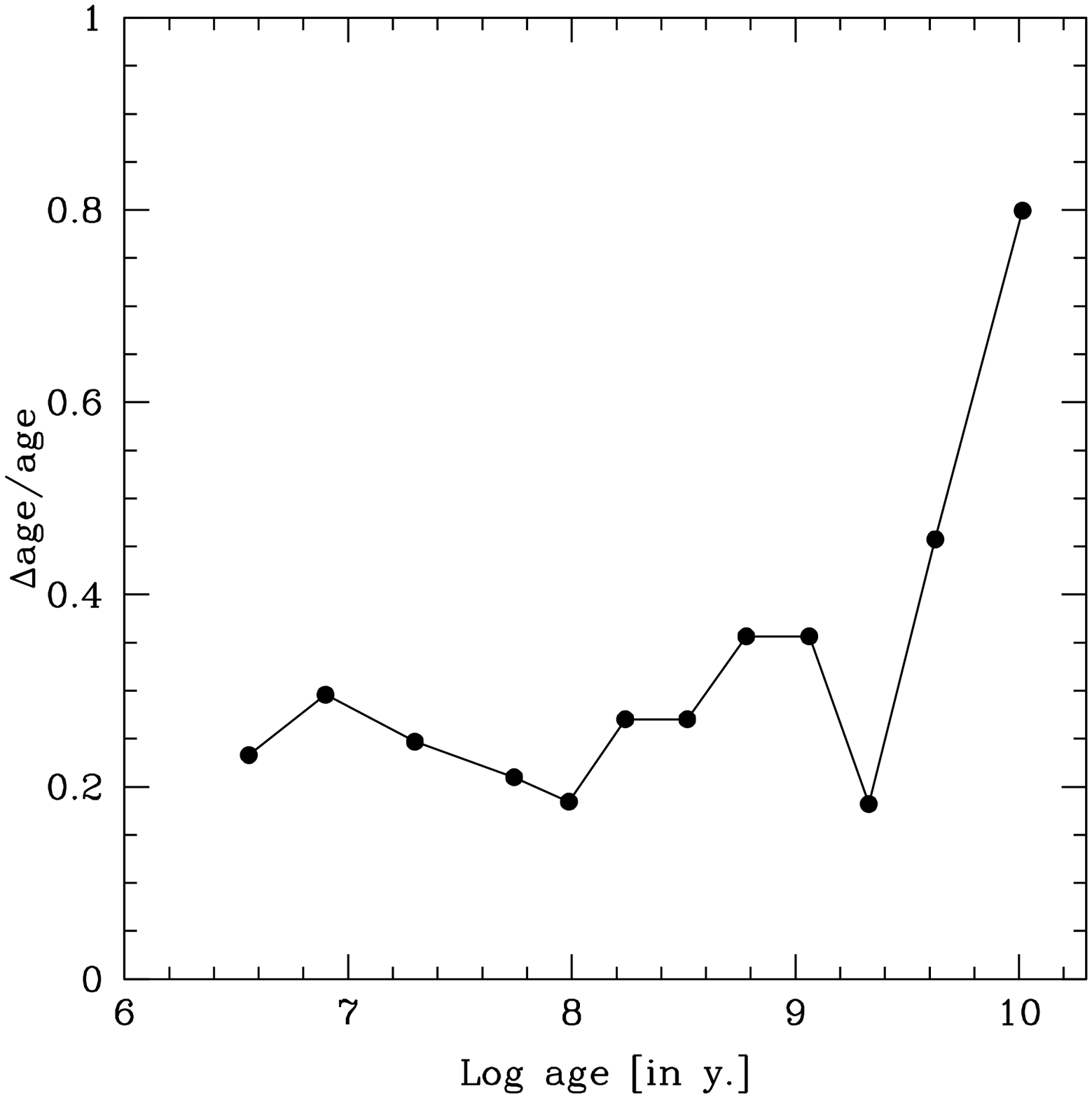}
\caption{{\it Left panels}: bolometric magnitude at the turn off versus the age for non-rotating stellar models at different metallicities.
{\it Right panel}: dispersion of the ages due to a change of the metallicity (see left panel for the list of models used).}\label{fig2}
\end{figure}

In Fig.~\ref{fig2} (left panel) relations between the bolometric magnitude at the turn off and the age
are shown for various metallicities (see references on the figure). When the metallicity increases, a given magnitude is achieved by a higher initial stellar mass since for the Z range considered here the luminosity of the tracks for a given initial mass are decreased. Thus smaller ages are obtained at higher Z for a given value of the turn off magnitude. In Fig.~\ref{fig2} (right panel) we have plotted the age dispersion which would result from using tracks for different initial metallicities in the range between Z=0.001 and Z=0.040 (i.e. for [Fe/H] between about -1.3 and 0.3). We see that the dispersion is higher than the one resulting from different stellar grids with similar metallicities. It amounts
to about 30\% for ages below about 3 Gyr. It increases a lot for greater values of the ages, reflecting the increased sensitivity of the luminosity on Z in the low mass range. This demonstrates that precise determination of the age needs relatively precise determination of the metallicity.

Dotter et al. (2007) have recently studied the impact of individual changes in the abundances of some elements. They computed models for stars with masses between 0.5 and 3 M$_\odot$ enhancing the abundance of one element keeping X, Y and Z constant. Of course enhancing one element at constant Z must be done at the expense of all other elements. This work shows that the elements which have the most important effects are oxygen and iron. Varying their abundances by a factor two
produces changes of the MS lifetime for a given initial mass star of 15\% (decrease for O and increase for Fe, see their
Fig.~13). 
Let us note that the mass fraction of helium also has a big impact on the structure and the evolution of stars. For instance, models by Claret (1997) or Bono et al. (2000)  have been computed with different He mass fractions. We refer the reader to these works for more details on that question (see also the paper by Decressin et al. in the present volume).

\section{Treatment of convection}

The treatment of convection and more generally of all turbulent processes remains one of the biggest difficulty in stellar modeling. Depending on the criterion chosen for the set up of the convective instabilities (Schwarzschild or Ledoux criterion), on the efficiency of semiconvection, on the amplitude of the overshooting effects, the quantity of fuel, the luminosity and therefore the MS lifetime can vary a lot.  For instance, a moderate overshoot (0.2 H$_p$) at the border of the Schwarzschild convective core associates an age at turn off luminosities equal to  2 [log L/L$_\odot$] (turn off mass around 3 M$_\odot$), 1 (1.6 M$_\odot$) and 0.5 (1.3 M$_\odot$) which are respectively 40\%, a factor 2.5 and a factor 3.2 greater than the associate lifetimes obtained from models without overshoot. One sees that compared to the age dispersion arising from different grids of models (but with similar physical ingredients) or from grids at various metallicities, the age dispersion due to overshoot is much greater, especially in the low
mass range. 

It has to be noted that the luminosity at the turn off is particularly sensitive to the amount of the overshoot. This property leads many authors to use the width of the observed MS to calibrate the overshhoot (see for instance Maeder \& Meynet 1989). Note that the temperature of the MS termination is much less sensitive to overshoot. This is due to the fact that overshooting increases the age but also pushes the MS termination to the red in such a way that the log T$_{\rm eff}$ at the MS termination versus age relation is nearly not changed (see Figs~17 \& 18 in Maeder \& Meynet 1989). However, if
the overshoot does not change the age assigned to a given turn off effective temperature, it does affect the initial mass associated to it!

Asteroseismology will likely help to resolve the question of the size of the convective core in massive stars. First results are presented in Aerts (2008): in five B-type star, the size of the convective core has been deduced from asteroseismology. An overshoot parameter between 0.10 and 0.44 H$_p$ has been found. 

In small mass stars, non-adiabatic convection occurs in the outer layers. The extent of the convective zone is  governed by the choice of the mixing length parameter $\alpha$, the value of it being fixed by calibrating solar models and/or from the position of the red giant branch in the HR diagram.
Of course (as for the overshooting parameter) $\alpha$ is not a fundamental constant of nature and it may vary with the mass, the metallicity, and even the evolutionary phase. Thus it would be advantageous to have
a mean of constraining this quantity in other objects. Asteroseismology may be the tool to do that. Eggenberger et al. (2008) show that the observed constraints of 70 Oph A can be reproduced by two sets of very different models: one with an initial helium value of 0.266, a mixing length
parameter equal to 1.7998 (obtained from a solar model) and an age of 6.2 Gyr, another with
an initial helium value of 0.240, a mixing length
parameter equal to 2.2497 (previous value multiplied by 1.25) and an age of 10.5 Gyr. Thus we see that the uncertainties on $\alpha$ may have a deep impact on the age determination. While these two models show the same mean large separations, the mean small separation of the model 
with the higher initial helium abundance is significantly larger (3$\mu$Hz) than the one of the model with the lower initial helium abundance. Thus having a precise observed value of the mean small separation will allow to obtain an independent determination of the age, of the mixing-length parameter and of the helium abundance.

\section{Rotation}

Rotation does affect all the outputs of the stellar models and in particuliar the age associated to a given initial mass star (Heger \& Langer 2000; Meynet \& Maeder 2000).
By inducing internal mixing, rotation modifies both the total quantity of 
fuel available and the luminosity. The amplitude of the changes depends
on the nature of the instabilities (induced by rotation) which are considered in the model. The two most important instabilities are shear instabilities and meridional currents. They both transport chemical species and angular momentum.
They are much less efficient in transporting energy since their timescale is in general longer than
the thermal diffusion timescale (see Zahn 1992). 

Before discussing the implications of rotation on the age determinations, let us say a few words about the observations which can be used to constraint rotational mixing.
A rotating star is predicted to present some nitrogen surface enrichment already during the main sequence. The amplitude of the nitrogen enrichment at the surface depends on the initial mass (increases with the mass), the age (increases with the age) and the initial rotational velocity.
This is correct as long as we consider stars with a given initial composition (rotational mixing is more efficient at low Z) and whose evolution is not affected by a close binary companion.
Thus we see that the nitrogen surface abundance is at least a function of three parameters:
mass, age and velocity. To see a relation between N-enrichment and velocity, it is necessary
to use stars with different rotational velocities but having similar masses and ages. 



\begin{figure}
\includegraphics[width=2.5in,height=2.5in]{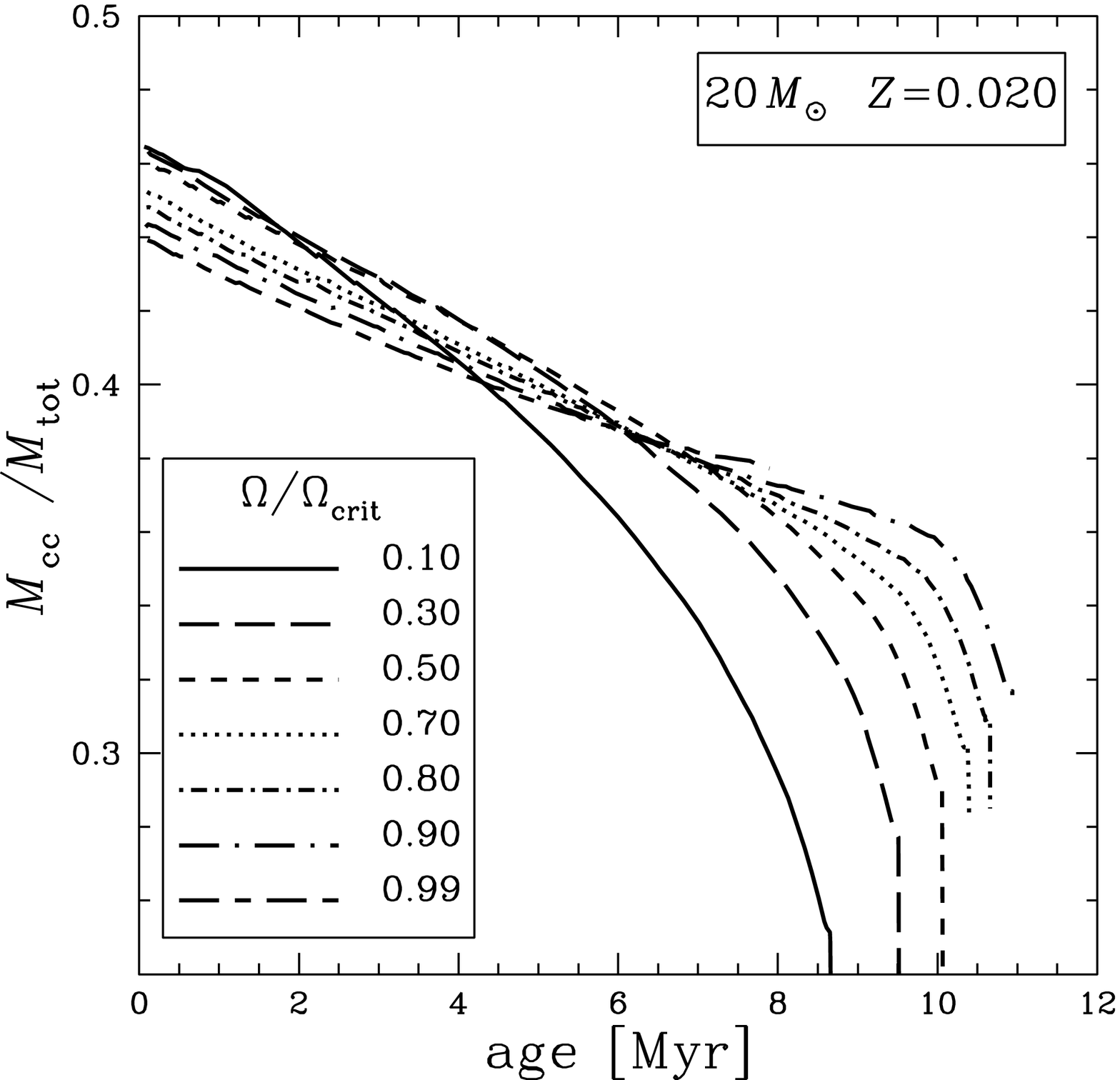}
\hfill
\includegraphics[width=2.6in,height=2.6in]{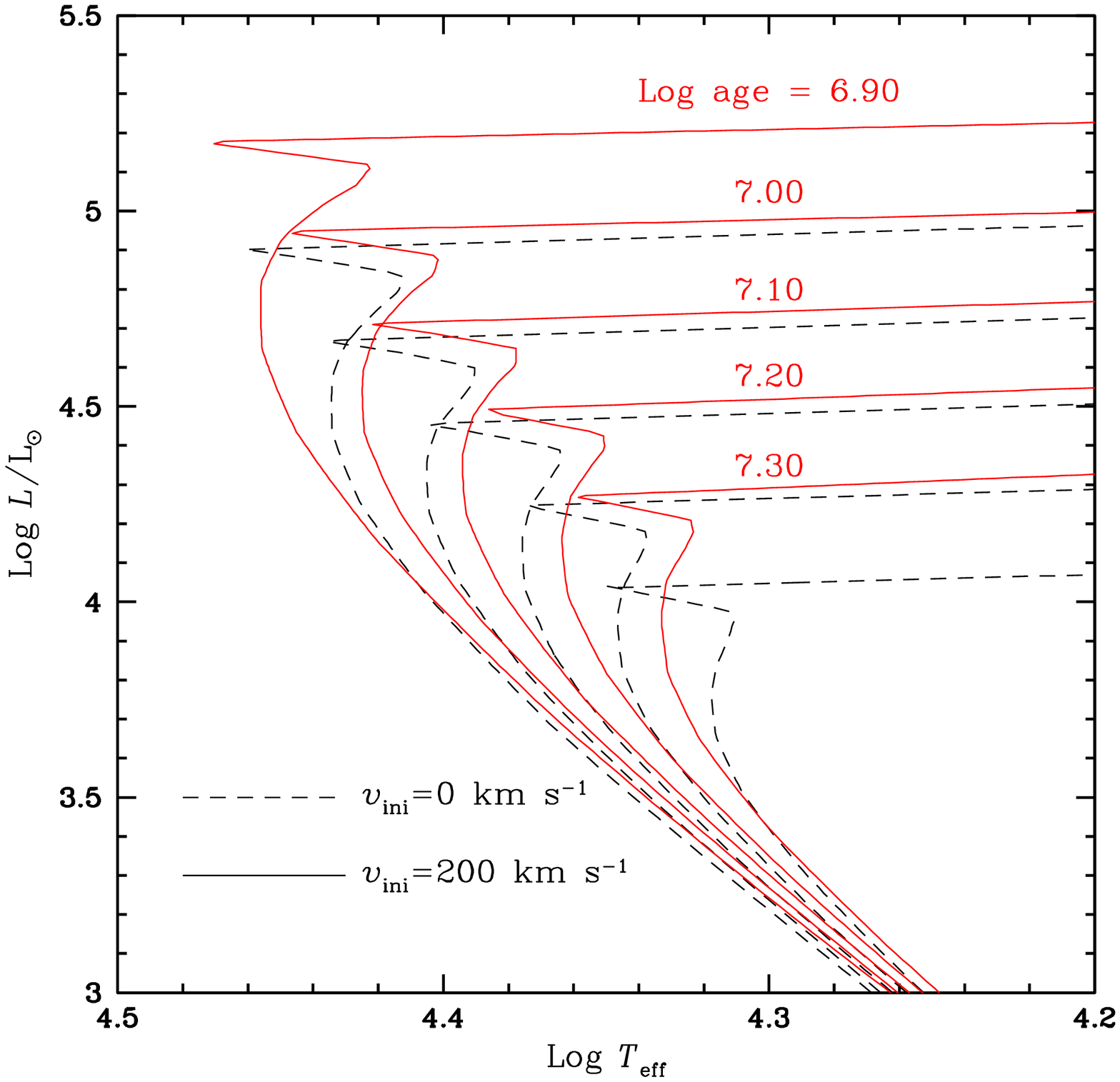}
\caption{{\it Left panels}: Evolution of the size of the convective core during MS, for the
      various values of rotational rate, in the 20 M$_\odot$ at standard
      metallicity. Figure taken from Ekstr\"om et al. (2008).
{\it Right panel}: Isochrones computed from stellar evolutionary tracks for the solar metallicity. 
The dashed and continuous lines correspond to the case of non--rotating and rotating 
stellar models respectively. In this last case, models have
an initial velocity $v_{\rm ini}$ of 200 km s$^{-1}$. The logarithms
of the ages (in years) label the isochrones computed from the models with rotation. Figure Taken from Meynet \& Maeder (2000)}\label{fig4}
\end{figure}

When data samples limited in mass and ages are used, a very nice correlation is found between
the surface N-enrichment and $\upsilon\sin i$ (see Figs. 3 and 4 in Maeder et al. 2008), 
supporting a N enrichment depending on rotational velocities. Stars beyond the end of the MS phase
do not obey to  such a relation, because their velocities 
converge toward low values (see Fig.~12 by Meynet \& Maeder 2000). 
A fraction, which we estimate to be   $\sim20$ \%
of the stars, may escape from the relation as a result of binary evolution, either by tidal mixing or mass transfer.

Rotation modifies the size of the convective core (see Fig.~\ref{fig4} left panel). First the size of the convective core is decreased when rotation is accounted for. Indeed due to the action of the centrifugal acceleration the star behaves as a star with a lower gravity or as a star of lower initial mass. Then, when evolution proceeds, rotational diffusion will supply the core with hydrogen and the radiative envelope with helium. The result
will be an increase of the luminosity and of the size of the convective core. 
The MS lifetimes are increased by rotation. During the MS phase, toward the turn off, the tracks for a given mass in the HR diagram
become more luminous  and extend further in the cool part. In that respect
the effects of rotation are somewhat similar to that of an overshoot (but of course the effects
of rotation cannot be modelized by adding an overshoot. For instance, as indicated above, rotation modifies the surface abundances, while overshooting does not). A given
width of the MS can be obtained by a combination of rotation velocity and overshooting parameter
(see Talon et al. 1997). For instance
in the temperature range between 4.1 and 4.3 (i.e. in the mass range between about 5 and 20 M$_\odot$)
a similar MS width can be obtained either by non-rotating models with an overshooting parameter of 0.2 or rotating models with an average rotational velocity on the MS around 100 km s$^{-1}$ and
an overshoot of 0.1 (estimates based on the models discussed in Ekstr\"om et al. 2008).
Rotation and overshooting thus concur to make the MS band wider. Calibrating the overshooting parameter in non-rotating stellar models by adjusting the MS band width thus overestimates it. How can we disentangle the two processes? Comparisons of the core size, obtained by asteroseismology, in stars of similar mass,
metallicity and age but having different rotational velocity will give precious constraints. Interestingly also the asteroseismic signature of an overshoot producing a sharp chemical gradient
at the border of the convective core is different from the one left by a smoother gradient produced
by rotational mixing (see Montalban et al. 2008).
Another approach would be to calibrate the overshooting parameter using stars massive enough to have a convective core but presenting a velocity distribution biased toward slow rotators.
Typically early type F-stars would fit the conditions (masses between 1.3 and 1.5 M$_\odot$), although
recent rotational velocity determinations (see Royer et al. 2007) show that the mean velocity of these stars is not as low as previously thought. According to Royer et al. (2007), the mean velocity of these stars is in the range of 150 km/s instead of 80-100 km/s.
Thus the effects of rotation are probably not completely absent (although much less developed than in massive stars). Another difficulty in this mass range comes from the fact that probably
the overshooting parameter varies with the mass as has been found by many authors using non-rotating models (see e.g. Girardi et al. 2000; Demarque et al. 2004). Thus the problem is not so easily resolved and its solution should probably wait more
stringent constraints coming from asteroseismology (see the talk by Vauclair in the present volume).

To have an idea of the effects of rotation on the age determination of young clusters, Fig.~\ref{fig4}, right panel,  compares isochrones obtained from non-rotating and rotating models. The rotating models have a rotational velocity of $\sim$150 km/s on average during the MS phase. We see that the age determined from rotating models are $\sim$25\% higher than those obtained from non-rotating ones (Meynet \& Maeder 2000). 

When the rotation is high enough significant changes of the shape of the star are expected.
Typically for a 20 M$_\odot$ star, when $\Omega/\Omega_c > 0.7$ (average velocity on the MS
superior to 280 km/s), then the equatorial radius is more than 10\% larger than the polar one.
The von Zeipel theorem (1924) implies then that the radiative flux in the polar region is higher than in the equatorial one. This can be translated into a change of the effective temperature with the colatitude. 

A consequence is that fast rotators will be characterized by different values of the luminosity and
of the effective temperature depending on the angle of view. 
For a fast rotator (initial velocities on the ZAMS such that $\Omega/\Omega_{\rm crit}\sim 0.90$) the dispersion in age due to the dispersion of the angle of view can amount to about 40\%  (using the luminosity at the turn off for determining the age, higher when the star is seen pole on).

In a cluster we have a distribution of rotational velocities and of directions of rotational axis. 
Thus we expect that the above effects will induce some dispersion in the position of the stars in the HR diagram around the turn off. 
Such effects are probably to be accounted for when determining the age of clusters containing significant proportions of Be-stars
(i.e. B-type stars with emission lines, the emission is due to the presence of an expanding
equatorial disk whose origin is commonly attributed to the very fast rotation of the star). This is the case for instance in NGC 330, a cluster of the SMC,
(see e.g. Keller et al. 2000), in which  40\%  of the MS stars which are two magnitudes below the turn off are Be stars. Note that in the bin of half a magnitude width just below the turn-off, 80\% of stars are Be stars!

\section{Microscopic diffusion}

Microscopic diffusion appears as soon as a fluid is out thermodynamic equilibrium.
Any gradient of temperature, density, external forces like gravity or radiative forces will
induce a diffusion velocity. In most cases, the diffusion velocities are much weaker than other velocities in the medium, therefore microscopic diffusion can only have a significant effect 
in very stable media.

In non-rotating solar mass models,
microscopic diffusion tends to reduce the MS lifetime by about 10\% (model computed by
P. Eggenberger for the present paper). The lifetime is reduced mainly because helium diffuses in the H-burning region, hydrogen is pushed outside from the core by mass conservation. Diffusion thus decreases the amount of available fuel for the star and thus the lifetime.
When a small rotation is accounted for (initial velocity of about 50 km s$^{-1}$), the lifetime reach
back nearly the same value as the model without diffusion and without rotation. This comes from the fact that the two effects nearly compensates, diffusion decreasing the quantity of fuel in the core and rotation
increasing it.
Eggenberger et al. (2008) find that accounting or not for the effect of diffusion changes the age
estimate of the binary system 70 Oph AB (masses of about 0.9 and 0.7 M$_\odot$) by about 1 Gyr, which makes a relative change of about 
14-16\%.

\begin{table}
\caption{Tentative estimates of the dispersion of ages obtained from models of different authors or computed with different physical ingredients.}
\begin{center}\scriptsize
\begin{tabular}{lccl}
 & & & \\
Cause                    & Range of ages & Relative error    & Remarks                            \\ 
 & & & \\
Origin of stellar models  & all           & 10-30\%           & for similar input physics          \\
metallicity              & 6.5 - 9.5     & 20-35\%           & for $-1.3 <$[Fe/H]$<0.3$           \\
                         & 9.5 - 10.0    & 35-80\%           & for $-1.3 <$[Fe/H]$<0.3$           \\
                         &               & 15\%              & changing O or Fe by a factor 2     \\
overshooting             & 6.5 - 9.5     & 40-320\%          & for 0 $<$d$_{ov}$/H$_p\le$0.20     \\
mixing-length            & $>$ 9.5          & 70\%              & based on the study of 70 Oph A     \\
Rotation                 & $<$ 9.5         & 25\%              & for normal rotational velocities   \\
                         & $<$ 9.5         & 40\%              & mostly due to angle of view (fast rotators only)       \\ 
Microscopic diffusion    & $>$ 9.5          & 15\%              & based on the study of 70 Oph AB    \\
Mass loss                & $<$ 6.5          & 25\%              & difference between 1 X and 2 X Mdot\\
 & & & \\
\end{tabular}
\end{center}
\label{t1}
\end{table}

\section{Mass loss}

The effects of mass loss during the MS phase begins to become significative for stars
more massive than about 30 M$_\odot$ at solar metallicity (that means ages inferior to about 5 Myr).
Mass loss makes a star of a given mass to follow an evolution in the HR diagram similar to that
of a star with a lower initial mass.
This tends thus to reduce the luminosity and to increase the MS lifetime. It may also make very massive initial stars to enter into the WR phase already during the MS phase. The MS termination then occurs at younger ages and at lower luminosities than in the case without mass loss.
Mass loss significantly affects the Magnitude at the turn off versus age relation only for ages inferior to about 3.3 million years. High mass loss reduces the age corresponding to a given value of the magnitude at the turn off. For $M_{bol}$ at the turn off =-10.8 for instance the difference of ages between the one deduced from normal and enhanced mass loss rate model amounts to 0.73 Myr, passing from 3.55 Myr (normal mass loss, Schaller et al. 1992) to 2.62 Myr (enhanced mass loss, Meynet et al. 1994).

Depending on the intensity of mass loss the evolution as a function of time of the number fraction of WR to O-type star after a starburst episode is different. Thus the age determination based on the deduction of this number fraction estimated from emission line ratios is affected by uncertainties of the mass loss rates. For instance a young starburst presenting a WR/O number ratio of 0.20  at Z=0.02 may have an age of 4.25 or of 3.25 Myr
when respectively standard or enhanced mass loss rates are used (Meynet 1995). In case of starbursts other factors
have an impact on the age determination as the timescale for the star formation episode (in the example above, we supposed an instantaneous episode), or the slope of the IMF .

\section{Conclusions}
 
From Table 2 below we can estimate the accuracy which would be needed on various physical ingredients of the models in order to reach let'say a 10\% accuracy on the age. We see that a precision better than 0.12 dex in [Fe/H] is needed
to assure a 10\% precision on the age over the whole range of ages, and that individual abundances of oxygen and of iron should be known with an accuracy better than a factor 2. The size of the convective core should be known with a very high precision especially in the lower mass range. As explained above this question is intimately related to the effects of rotation. The value of $l/H_p$ should be known with an accuracy better that 5-10\%, the mass loss with a precision better than about 20\%.
 
As usual improvements will be brought by progresses in observations and in our physical understanding of turbulence in stellar conditions. In the observational context, asteroseismology
will play a key role. This technic allows to probe the stellar interiors and provide, in addition to the classical  constraints as e.g. gravities and effective temperatures, new  observables, making much smaller the possible range of values for the free parameters. From the point of view of theory,
multi-dimensional computations of hydrodynamic processes like convection will provide more thoughtful recipes for accounting for the effects of turbulence in stellar interiors.

The list of effects in Table 2 is not exhaustive: what are
the effects of magnetic fields, of internal gravity waves, of tidal mixing in close binaries,
of accretion during the pre-MS phase of massive stars? All these effects still need
to be studied.

\end{document}